\documentclass[aps,prl,twocolumn,unsortedaddress,showpacs]{revtex4}%
\usepackage{graphicx}
\usepackage{epsfig}
\usepackage{amsmath}
\usepackage{amsfonts}
\usepackage{amssymb}

\newcommand{\btau}{\mbox{\boldmath$\tau$}}
\newcommand{\bepsilon}{\mbox{\boldmath$\epsilon$}}

\newcommand{\bS}{\mbox{\boldmath$S$}}
\newcommand{\bQ}{\mbox{\boldmath\small$Q$}}
\newcommand{\bq}{\mbox{\boldmath\small$q$}}

\newcommand{\bR}{\mbox{\boldmath\small$R$}}

\newcommand{\ba}{\mbox{\boldmath$a$}}

\newcommand{\br}{\mbox{\boldmath$r$}}

\newcommand{\jlarge}{\mbox{\large$j$}}
\newcommand{\ii}{\mbox{$\rm i$}}

\begin{document}

\title{Spiral order induced by distortion in a frustrated square-lattice antiferromagnet.}

\author{I.~A.~Zaliznyak}
\affiliation{Brookhaven National Laboratory, Upton, New York 11973-5000 USA.}

\begin{abstract}

In a strongly frustrated square-lattice antiferromagnet with diagonal coupling
$J'$, for $\alpha = J/(2J') \lesssim 1$, an incommensurate spiral state with
propagation vector $\tilde{\bQ} = (\pi \pm \delta, \pi \pm \delta)$ near $(\pi,
\pi)$ competes closely with the N\'{e}el collinear antiferromagnetic ground
state. For classical Heisenberg spins the energy of the spiral state can be
lowered as it adapts to a distortion of the crystal lattice. As a result, a weak
superstructural modulation such as exists in doped cuprates might stabilize an
incommensurate spiral phase for some range of the parameter $\alpha$ close to
$1$.

\end{abstract}


\maketitle

An interplay between small distortion of the crystal lattice and the magnetic
properties of the material is currently a subject of intense research. One
problem which supplies strong motivation for such studies is that of stripe
order in the lightly doped high-T$_c$ cuprates La$_{2-x}$Sr$_x$CuO$_{4+y}$
(LSCO) and in related nickelates \cite{TranquadaWakimotoLee,Tranquada1995}.
These phases are always associated with a weak superstructural distortion of the
original ``stacked square lattice'' structure of the un-doped parent material.
Incommensurate magnetism in these compounds is usually interpreted in terms of a
segregation of the doped charges into lines which separate the antiferromagnetic
domains (``stripes'') characteristic of the un-doped material. Although
modulation of the crystal structure which is induced by charge-stripe
segregation is often too small to be observed in experiment
\cite{TranquadaWakimotoLee}, it is clear that essential result of the stripe
order for the spin system of cuprates is a periodic modulation of the exchange
coupling in the Heisenberg spin Hamiltonian which describes their magnetic
properties \cite{Coldea2001}. So far, though, only the simplest ``average''
consequence of the stripe superstructure, in the form of the effective weakening
of exchange coupling in the direction perpendicular to the stripes, has been
considered \cite{CastroNetoHone}. A similar problem, of an interplay between the
spin order and the cooperative Jahn-Teller distortion accompanying the charge
order, arises in the context of the charge-ordered phases in doped manganites
\cite{Orenstein2000}.

Because the low-energy magnetic properties of layered LSCO cuprates are believed
to be adequately described by the two-dimensional (2D) Heisenberg spin
Hamiltonian, this model has recently become a focus of intense research. Special
attention was devoted to the frustrated square lattice, where in addition to the
nearest-neighbor exchange interaction, $J > 0$, there is a diagonal coupling,
$J' > 0$, such that $\alpha = \frac{J}{2J'}$ is close to 1. It was originally
motivated by the predictions that non-N\'{e}el resonating valence bond states
\cite{Anderson1987,Kivelson1987} and quantum-critical behavior
\cite{Chakravarty1988} associated with the $T = 0$ order-disorder phase
transitions which may occur in this case might be important for the physics of
the superconductivity in cuprates.

Despite RVB spin-liquid state and quantum criticality are strongly predicated
upon the quantum nature of the spins (S=1/2 in cuprates), a semiclassical
spin-wave theory appears to provide a surprisingly good guidance to the behavior
of the frustrated square-lattice antiferromagnet (FSLA)
\cite{ChandraDoucout,Dagotto1989,Sushkov,Zhitomirsky2000,Lieb1999,Singh1999,Tchernyshyov2003}.
Perhaps, this is because the phenomenon of frustration mainly rests on the
ground state degeneracy which exists for classical, as well as for quantum
spins. In fact, existence of the spin-liquid phase possibly related to the RVB
state in the FSLA for the range of the parameter $\alpha$ around $\alpha = 1$
was first conjectured in Ref. \onlinecite{ChandraDoucout} on the basis of the
conventional spin-wave calculation to the order 1/S. This suggestion was then
supported by the field-theory methods \cite{Sachdev}, numerical calculations
\cite{Dagotto1989,Sushkov,Zhitomirsky2000} and other studies
\cite{Lieb1999,Singh1999,Tchernyshyov2003}. It was established that a disordered
phase, whose nature is still controversial, is realized for $0.8 \lesssim \alpha
\lesssim 1.1$. Although these studies were essentially aimed at understanding
the physics of doped LSCO and related materials, the lattice modulation was
generally ignored. One reason for this is that, traditionally, the lattice
distortion in a spin system is treated by switching to a larger unit cell, with
multiple different spin sites. This approach is not viable for the long-periodic
superstructures, and is not possible for the charge-ordered states with
incommensurate modulation.

In contrast with the previous studies, present paper addresses the consequences
of the superstructural lattice distortion for the ground state of the 2D
Heisenberg spin Hamiltonian with classical spins, \emph{ie} essentially presents
an ``unrealistic'' mean-field (MF) treatment of the realistic spin model.
Although MF results are subject to significant quantum corrections, especially
for small spins S$\lesssim 1$, they nevertheless provide useful guidance about
the hierarchy of the competing ground states (GS) in the system. In fact, the MF
ground state very often survives account for quantum and thermal fluctuations,
as it does for the un-frustrated 2D antiferromagnet.

Main finding of this paper is that a weak superstructural modulation of the
crystal lattice in the FSLA may stabilize an incommensurate spin-spiral ground
state with the propagation vector $\tilde{\bQ} = (\pi \pm \delta, \pi \pm
\delta)$ close to $(\pi, \pi)$ for $\alpha \leq 1$.
%Spiral state appears as a local minimum along the $(q,q)$ direction
%of the MF Heisenberg exchange energy for classical spins
Although in the absence of a structural modulation the energy of the spiral
states is higher than that of the collinear N\'{e}el states illustrated in Fig.
1(a) (except for $\alpha = 1$), they are in close competition for $\alpha$ near
1. While spin spiral is usually ignored in the analysis of the possible phases
in FSLA, in presence of a superlattice modulation it might actually win the
competition for some range of $\alpha \lesssim 1$. Here this is shown explicitly
on the mean field level, by treating the effect of a \emph{small} but otherwise
quite \emph{arbitrary} lattice distortion, as a perturbation in the microscopic
classical-spin Heisenberg Hamiltonian.

Consider a system of $N$ equivalent spins on a square lattice, Fig. 1 (a),
coupled by Heisenberg exchange interaction, ${\cal H} = \sum_{i,j} J_{ij}\,
\left(\bS_i \bS_j\right)$. While only coupling between the nearest neighbors
along the side ($J$) and along diagonal ($J'$) will be of interest in this
paper, here $J_{ij}=J_{ji}$ parameterize a general exchange coupling between the
spins at arbitrary lattice sites $i$ and $j$. In the absence of a distortion the
MF classical ground state is a planar transverse spin spiral, $\bS_j = (S\cos
(\bQ \br_j), S\sin(\bQ \br_j),0)$,
\cite{YoshimoriVillainLyonsKaplanNagamiya,ZaliznyakZhitomirsky}. The ordering
wave vector $\bQ$ corresponds to the minimum of the lattice Fourier transform of
the exchange interaction, $J_{\bq} = \sum_{\br_{ij}} J_{ij} \exp(-\ii \bq
\br_{ij})$, $\br_{ij} = \br_j - \br_i$. This GS is obtained by finding the
minimum-energy configuration for the Heisenberg Hamiltonian with classical spins
under the constraint that $\bS_j^2 = S^2$ for all sites $j$. In general case,
spontaneous symmetry breaking is defined by the two mutually perpendicular spin
vectors which determine the polarization of the spiral, \emph{ie} by the
Fourrier transform of the lattice spin distribution, $\bS_{\bQ} = \bS' + \ii
\bS''$. For collinear situations, such as ferro- or antiferromagnet,
corresponding to $\bQ = 0$ and, \emph{eg}, $\bQ = (\pi,\pi)$, respectively, only
a single vector is needed for the order parameter.

%***************************Figure 1*****************************************
\begin{figure}[t]
\vspace{-0.in}
\begin{center}
\label{modulated_square}%
\includegraphics[width=2.in, ]
{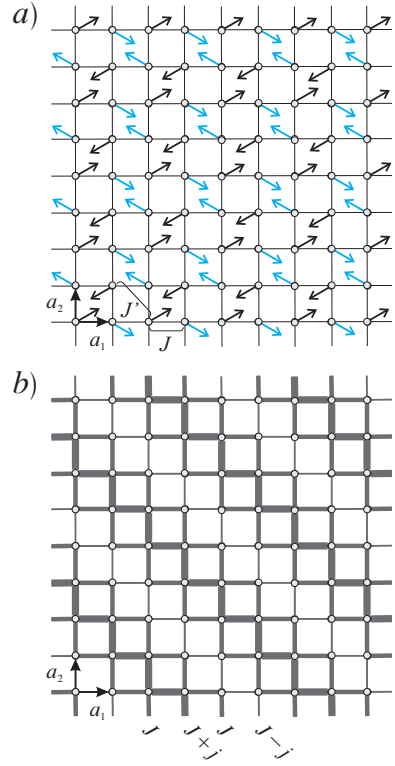}%
\caption{(a) Frustrated square lattice with diagonal coupling $J/(2J') < 1$;
continuous degeneracy of the MF ground state corresponds to an arbitrary angle
between the two antiferromagnetic sublattices. (b) ``Stripes'' on the square
lattice with diagonal modulation, $Q_c = (\frac{2\pi}{n}, \frac{2\pi}{n})$,
$n=4$ case is shown. }
\end{center}
\vspace{-.3in}
\end{figure}
%***************************Figure 1*****************************************

A slight distortion of the crystal structure which is characterized by the
appearance of the additional, weak supperlattice Bragg reflections at
wavevectors $\pm \bQ_c$, corresponds to a small harmonic modulation of the ionic
positions, $(\br_j)' = \br_j + \bepsilon_1 \cos(\bQ_c \br_j) + \bepsilon_2
\sin(\bQ_c \br_j)$. In most general case, this results in a harmonic modulation
of the exchange coupling. It has either the same wavevector $\bQ_c$, if it
appears as a first-order correction to $J_{ij}$ in small parameter $\epsilon
\sim \left( \frac{\epsilon_{1,2} }{r_{ij}} \right) \ll 1$, or the wavevector
$2\bQ_c$, if it appears only in the second order, $\sim \epsilon^2$,
\cite{Zaliznyak2003}. There is also a second-order correction to the bond
energy, $\tilde{J_{ij}} = J_{ij} + \delta J_{ij}$. The spin Hamiltonian becomes,
\begin{equation}
\label{H_mod}%
{\cal H} = \sum_{i,j} \left( J_{ij} + \jlarge_{ij} {\rm e} ^{\ii  \bQ_c
\bR_{ij}} + \jlarge^*_{ij} {\rm e} ^{-\ii \bQ_c \bR_{ij}} \right) \bS_i \bS_j
\;.
\end{equation}
where the tildes were omitted, and the complex $\jlarge_{ij} = \jlarge'_{ij} +
\ii \jlarge''_{ij}$ was introduced. While without distortion $J_{ij}$ would
satisfy all symmetries of the lattice, exchange constants in Eq. (\ref{H_mod})
possess only those symmetries of the un-distorted lattice which preserve
$\bQ{_c}$ and the distortion polarizations $\bepsilon_1,\bepsilon_2$ (this
includes all translations).

The modulated-exchange terms allow um-klapp processes which couple $\bS_{\bq}$
and $\bS_{\bq \pm \bQ_c}$ in the spin Hamiltonian, and couple these
Fourrier-components in the equations expressing the conditional minimum of the
classical exchange energy. As a result, additional Fourrier harmonics, at
wavevectors $\bQ + n \bQ_c$, $n = \pm 1, \pm 2,...$, appear in the GS spin
structure. It has the form of expansion,
\begin{equation*}
\label{Sq_expansion}%
\bS_{\bq} = \sum_{n} \bS_{\bQ + n\bQ_c} \delta_{\bq, \bQ + n\bQ_c} + \bS_{\bQ +
n\bQ_c}^* \delta_{-\bq, \bQ + n\bQ_c},
\end{equation*}
where $S_{\bQ + n\bQ_c} \sim O(\epsilon^{|n|})$. This corresponds to a bunched
spiral
\cite{YoshimoriVillainLyonsKaplanNagamiya,ZaliznyakZhitomirsky,Zaliznyak2003}.
Based on very general exchange symmetry arguments
\cite{ZaliznyakZhitomirsky,Zaliznyak2003,AndreevMarchenko}, in the absence of
any additional symmetry breaking, the perturbing terms have to be proportional
to the non-perturbed order parameter. As a result, the leading new
Fourrier-components, $\bS_{\bQ \pm \bQ_c}$, are,
\begin{align}
\label{S_Q_solution_1}%
& \bS_{\bQ + \bQ_c} = \left[ \frac{ \jlarge_{\bQ - \frac{1}{2} \bQ_c} -
\jlarge_{\bQ + \frac{1}{2} \bQ_c} }{ \chi_\perp ({\bQ_c}) \omega^2_{\bQ_c}
S^{-2} } \right] \bS_{\bQ} + O(\epsilon^3), \\
\label{S_Q_solution_2}%
& \bS_{\bQ - \bQ_c} = \left[ \frac{ - \jlarge^*_{\bQ - \frac{1}{2} \bQ_c} +
\jlarge^*_{\bQ + \frac{1}{2} \bQ_c} }{ \chi_\perp ({\bQ_c}) \omega^2_{\bQ_c}
S^{-2} } \right] \bS_{\bQ} + O(\epsilon^3).
\end{align}
Here $\omega_{\bq} = S \sqrt{2(J_{\bq} - J_{\bQ})(J_{\bq + \bQ} + J_{\bq - \bQ}
- 2 J_{\bQ})}$ is the spin-wave spectrum in the initial, non-distorted,
single-$\bQ$ exchange spiral, $\chi_{\perp} (\bq) = \left[ 2(J_{\bq} -
J_{\bQ})\right]^{-1}$ is its transverse (perpendicular to the spin plane),
$\bq$-dependent staggered static spin susceptibility, and $\jlarge_{\bq} =
\sum_{\br_{ij}} \jlarge_{ij} \exp(-\ii \bq \br_{ij}) = \jlarge_{-\bq}$ is a
lattice Fourrier-transform of the modulated exchange term.
%\cite{JensenMackintosh}
Neglecting the $O(\epsilon^4)$ terms, the corrected ground state energy is
obtained in the second order of the perturbation theory,
\begin{align}
\label{E_GS_solution}%
\frac{E_{GS}}{N} = J_{\bQ} S^2 - \frac{ \left| \jlarge_{\bQ - \frac{1}{2} \bQ_c}
- \jlarge_{\bQ + \frac{1}{2} \bQ_c} \right|^2 S^4}{ \chi_\perp ({\bQ_c})
\omega^2_{\bQ_c}} ,
\end{align}
and, correspondingly, is, in general, lowered by the exchange modulation. This
occurs as a result of the appropriate adjustment (bunching) of the initial
single-$\bQ$ spiral spin structure, through appearance of the additional
Fourrier-harmonics, $\bS_{{\bQ} \pm \bQ_c }$. In addition, the pitch of the
primary spiral component, $\bS_{\bQ}$, may also change, $\bQ \rightarrow
\tilde{\bQ}$, because the spiral propagation vector, $\tilde{\bQ}$, is now
defined by the minimum of the corrected energy, Eq. (\ref{E_GS_solution}).

A singular situation occurs when the lattice modulation has the wavevector
$\bQ_c$ which is near the dispersion soft spot of the initial spiral, \emph{eg}
close to its Goldstone mode. In this case corrections (\ref{S_Q_solution_1}) -
(\ref{E_GS_solution}) diverge, and the perturbation approach fails, highlighting
the sensitivity of spin system to such modulations. In frustrated spin systems,
entire soft regions, such as the lines of soft modes, often appear due to the
accidental cancellation of the interactions. As a result, such systems must be
extremely sensitive to structural distortions. On the other hand, in many
important cases, such as a nearest-neighbor non-frustrated antiferromagnet,
$\bQ$ is a special symmetry point of $ \jlarge_{\bQ}$ ($ \jlarge_{\bQ} \sim
\epsilon J_{\bQ}$), and the correction term vanishes. Therefore, simple
structures, such as collinear antiferromagnets, are, in general, not sensitive
to small lattice modulations. In what follows, the singular cases will be
excluded from the consideration.

The results obtained above can now be applied to analyze the effect of lattice
modulation in a square-lattice antiferromagnet, which may be of direct relevance
for the charge-ordered phases in doped LSCO cuprates and related perovskites.
For definitiveness, consider the case of $n$-periodic diagonal modulation with
$\bQ_c = (\frac{2\pi}{n}, \frac{2\pi}{n})$, where $n = 2,3,4,...$, illustrated
in Fig. 1(b) (in LSCO the most stable superstructure occurs for $n=8$,
\cite{TranquadaWakimotoLee}). Without frustration, $\jlarge_{\bq} = \epsilon
J_{\bq}$, and, upon switching to $Q = \bq \frac{(\ba_1 + \ba_2)}{2}$ and $Q' =
\bq \frac{(\ba_1 - \ba_2)}{2}$, the problem is factorized and corrections are
essentially the same as for 1D chain. There is no change of the global minimum
of classical spin energy, so the nearest-neighbor antiferromagnetism is stable
with respect to the bond modulation.

In the frustrated case, $J_{\bq} = 4J \cos{Q} \cos{Q'} + 2J'( \cos{2Q} +
\cos{2Q'} ) $, and, if both side and diagonal bonds are modulated,
$\jlarge_{\bq} = 4\jlarge \cos{Q} \cos{Q'} + 2\jlarge'( \cos{2Q} + \cos{2Q'} )$.
Upon account for distortion the GS energy is, %becomes,
\begin{align}
\label{E_GS_1b}%
\frac{E}{N S^2} = J_{\bQ} + \frac{ \sin^2 Q \left| \jlarge \cos{Q'} + 2\jlarge'
\cos {\frac{\pi}{n}} \cos{Q} \right|^2 }{J \cos{Q} \cos{Q'} + 2 J'
\cos^2{\frac{\pi}{n}} \cos{2Q}}.
\end{align}

In the absence of bond modulation, the ground state is determined by the
hierarchy of the local minima of $J_{\bQ}$, which only depends on $\alpha =
\frac{J}{2J'}$. For weak frustration, $\alpha > 1$, the global minimum is that
with $\sin{Q} = \sin{Q'} = 0$. It corresponds to the conventional, collinear
N\'{e}el antiferromagnetic order with a single propagation vector $\bQ =
(\pi,\pi)$, and the ground state energy $\frac{1}{N}E_{(\pi,\pi)} = -4J S^2 (1 -
\frac{1}{2\alpha})$. Although there are four equivalent $\bQ$-points in the
Brillouin zone (BZ), $(\pm \pi, \pm \pi)$, $(\mp \pi, \pm \pi)$, which restore
the lattice $C_4$ rotational symmetry, they are related through addition of the
appropriate reciprocal lattice vectors $\btau$, so there is no true GS
degeneracy in $\bQ$-space. The only degeneracy is the GS rotational symmetry in
spin space, which is left of the $O(3)$ symmetry of the Heisenberg spin
Hamiltonian.

%$\alpha \leq 1$, additional GS degeneracy appears on the MF level. For
For strong frustration, $\alpha < 1$, there are two non-equivalent lowest-energy
minima of $J_{\bQ}$, they satisfy $\cos{Q} = \cos{Q'} = 0$ and have the GS
energy $\frac{1}{N}E_{(\pi,0)} = -4J' S^2 = -4J S^2 \frac{1}{2\alpha}$. They
correspond to two pairs of equivalent $\bQ$-points in the BZ, $(\pm \pi, 0)$ and
$(0, \pm \pi)$, which represent the antiferromagnetic order propagating along
the $X$ and $Y$ axis, respectively. This double degeneracy in $\bQ$-space can be
used to construct a continuum of states which are the linear combinations of the
above two. This continuous GS degeneracy is usually described in terms of two
decoupled antiferromagnetic sublattices based on the diagonals of the original
square lattice, which is transparent for $J' \gg J$. Each sublattice has an
antiferromagnetic order, but there may be an arbitrary angle between the two,
because the mean field from one sublattice cancels on the sites of the other,
Fig. 1a. This continuous degeneracy is lifted by zero-point or thermal spin
fluctuations which prefer collinear arrangements of the two sublattices in the
GS. This phenomenon is known as ``order from disorder''
\cite{Shender1982,Henley1989}.

Although it is not the focus of this paper, an interesting situation occurs for
$\alpha = 1$, when, on the MF level, there is also a continuous GS degeneracy in
the $\bQ$-space. The minimum condition for $J_{\bQ}$ becomes $\cos{Q} =
\cos{Q'}$, and is satisfied for any spiral with the propagation vector $\bQ$
that belongs to the square with the vertices at $(\pm \pi, \pm \pi)$, $(\mp \pi,
\pm \pi)$. They all have the same energy, $\frac{1}{N}E_{\alpha=1} = - 2J S^2 =
-4J' S^2$. This continuous $\bQ$-space degeneracy is at the origin of the
spin-liquid phase conjectured in FSLA for $\alpha$ close to $\alpha = 1$
\cite{ChandraDoucout,Dagotto1989,Sushkov,Zhitomirsky2000,
Sachdev,Lieb1999,Singh1999,Tchernyshyov2003}.

What is important here, is that the spiral states with $\bQ \approx (\pi,\pi)$
are in close competition with the collinear states for $\alpha \lesssim 1$. In
particular, the spiral with the propagation vector defined from $ Q' = 0$, $
\cos{Q} = - \frac{J}{2J'}$, \emph{ie} $\bQ = \left( \cos^{-1}( -\frac{J}{2J'}),
\cos^{-1}( -\frac{J}{2J'}) \right)$, is a local minimum of $J_{\bQ}$ along the
diagonal, $(q,q)$, direction, whose energy in the absence of modulation is
$\frac{1}{N}E_{Q} = - 2\alpha J S^2$.
%(there is also a degenerate state with $\bQ$ at $90^\circ$, respecting the $Q
%\leftrightarrow Q'$ symmetry of the square lattice).
Except for $\alpha = 1$, though, the energy of this local extremum (and of all
other spiral states) is higher than that for the decoupled antiferromagnetic
sublattices, $E_{(\pi,0)}$, and for this reason they are usually ignored.
However, it is clear from the Eq. (\ref{E_GS_1b}) that, while the energy of the
antiferromagnetic states is insensitive to the bond modulation, the \emph{energy
of the spiral state can be lowered as it adapts to the lattice distortion}!
Therefore, at least on the MF level, a spiral may become the lowest energy state
(\emph{ie} the ground state) for some range of the parameter $\alpha$ in the
vicinity of 1 (whose width is $\sim O( \epsilon^2)$). For a long-periodic
modulation, $Q_c \ll 1$, and for $\jlarge' = 0$, it is easy to find that the
spiral phase is stable for $1 - | \jlarge /J |^2 \lesssim \alpha < 1$. The
principal propagation vector of the spiral is obtained by minimizing Eq.
(\ref{E_GS_1b}).

While it would be interesting to study the modulated-exchange Hamiltonian
(\ref{H_mod}) for quantum spins and for the arbitrary values of $ |\jlarge_{ij}
/ J_{ij}|$, this is a formidable task which is beyond the scope of this paper.
Here a perturbative scheme is used to find the mean field ground state. It is
valid for classical spins, $S \gg 1$, and for small exchange modulation, $
|\jlarge_{ij} / J_{ij}| \sim \epsilon \ll 1$. Nevertheless, it provides an
important insight into behavior of the frustrated square-lattice
antiferromagnet. A finding that (by selecting the spiral order) exchange
modulation effectively destabilizes collinear N\'{e}el states preferred by the
fluctuations clearly supports the instability of the frustrated square-lattice
antiferromagnet with $J/(2J')$ close to 1 with respect to the bond-modulated
states, \cite{Lieb1999,Singh1999,Tchernyshyov2003}.

%Clearly, the results are most relevant for the localized-electron systems such
%as La$_{1.5}$Sr$_{0.5}$CoO$_4$ \cite{Zaliznyak2000}, where the itineracy effects
%are completely negligible. It provides a possible mechanism, which could
%explain, already on the mean field level, the spin incommensurability observed
%in these materials.
The essential results of this paper are summarized by Eqs.
(\ref{S_Q_solution_1}) - (\ref{E_GS_solution}). The main finding is that the
energy of the equal-spin transverse spiral state can be lowered by the exchange
modulation in the Heisenberg spin Hamiltonian. This happens as spiral adapts to
the modulation through appearance of the additional Fourrier-harmonics,
$\bS_{{\bQ} + n \bQ_c }$, $n = \pm 1, \pm 2, ...$ (bunching). As a result, in
frustrated square-lattice antiferromagnet with diagonal coupling $J'$, such,
that $\alpha = J/(2J')$ is close to 1, lattice modulation may open a region of
stability of the incommensurate spiral phase. This ``order by distortion''
phenomenon competes with ``order by disorder'', which prefers collinear
arrangements of two antiferromagnetic sublattices. Incommensurate spiral phase
with the propagation vector $\tilde{\bQ} = (\pi \pm \delta, \pi \pm \delta)$
close to $(\pi, \pi)$ wins for the range $O(\epsilon^2)$ of $\alpha$ around
$\alpha = 1$.

The arguments presented here provide plausible explanation for the
incommensurate spin-ordered phases, which are among the most interesting and
puzzling features observed in the doped perovskites, and may also be of direct
relevance for the doped LSCO materials.
%observed in La$_{1.5}$Sr$_{0.5}$CoO$_4$ \cite{Zaliznyak2000} and in a number of
For the Heisenberg spin Hamiltonian on square lattice in the absence of
distortion, one needs at least a third-neighbor coupling in order to stabilize
the MF spiral ground state.

It is a pleasure to thank F. Essler, S. Maslov and M. Zhitomirsky for
encouraging discussions, and acknowledge the DOE Contract \#DE-AC02-98CH10886.
%financial support under the DOE Contract \#DE-AC02-98CH10886 and by the Theory Institute at Brookhaven National Laboratory.


\begin{thebibliography}{99}                                                                                                %

\bibitem{TranquadaWakimotoLee}
 J.~M.~Tranquada {\it et~al.},
 %, J.~D.~Axe, N.~Ichikawa, A.R.~Moodenbaugh, Y.~Nakamura and S.~Uchida
 Phys.~Rev.Lett. {\bf 78}, 338(1997);
 Phys.~Rev.~B {\bf 59}, 14712 (1999).
% S.~Wakimoto {\it et~al.},
 %R.~J.~Birgeneau, M.~A.~Kastner, Y.~S.~Lee, R.~Erwin, P.~M.~Gehring,
 %S.~H.~Lee, M.~Fujta, K.~Yamada, Y.~Endoh, K.~Hirota, G.~Shirane
% Phys.~Rev.~B {\bf 61}, 3699 (2000);
% Y.~S.~Lee {\it et~al.},
 %R.~J.~Birgeneau, M.~A.~Kastner, Y.~Endoh, S.~Wakimoto, K.~Yamada,
 %R.~Erwin, S.~H.~Lee, G.~Shirane
% Phys.~Rev.~B {\bf 60}, 3643 (1999).

\bibitem{Tranquada1995}
 J.~M.~Tranquada \emph{et al},
%B.~J.~Sternlieb, J.~D.~Axe, N. Nakamura, S.~Uchida,
Nature {\bf 375}, 561, (1995).

\bibitem{Coldea2001}
 R. Coldea \emph{et al},
%S. M. Hayden, G. Aeppli, T. G. Perring, C. D. Frost, T. E. Mason, S.-W. Cheong, Z. Fisk
Phys. Rev. Lett. {\bf 86}, 5377 (2001).

\bibitem{CastroNetoHone}
 A.~H.~CastroNeto and D.~Hone, Phys. Rev. Lett. {\bf 76}, 2165 (1996).

\bibitem{Orenstein2000}
 J.~Orenstein and A.~J.~Millis, Science {\bf 288}, 468 (2000).

\bibitem{Anderson1987}
P.~W.~Anderson, Science {\bf 235}, 1196 (1987).

\bibitem{Kivelson1987}
S.~A.~Kivelson, D.~S.~Rokhsar, and J.~P.~Sethna, Phys. Rev. B {\bf 35}, 8865
(1987).

\bibitem{Chakravarty1988}
S.~Chakravarty, B.~I.~Halperin, and D.~R.~Nelson, Phys. Rev. Lett. {\bf 60},1057
(1988); Phys. Rev. B {\bf 39}, 2344 (1989).

\bibitem{ChandraDoucout}
P.~Chandra, B.~Doucot, Phys. Rev. B {\bf 38}, 9335 (1988).

\bibitem{Dagotto1989}
E.~Dagotto and A.~Moreo, Phys. Rev. Lett. {\bf 63}, 2148 (1989).

\bibitem{Sushkov}
O.~P.~Sushkov, J.~Oitmaa, Z.~Weihong, Phys. Rev. B {\bf 66}, 054401 (2002);
\emph{ibid} {\bf 63}, 104420 (2001); V.~N.~Kotov, O.~P.~Sushkov, \emph{ibid}
{\bf 61}, 11820 (2000).

\bibitem{Zhitomirsky2000}
M.~E.~Zhitomirsky, A.~Honecker, O.~A.~Petrenko, Phys. Rev. Lett. {\bf 85}, 3269
(2000).

\bibitem{Lieb1999}
E.~H. Lieb, P.~Schupp, Phys. Rev. Lett. {\bf 83}, 5362 (1999).

\bibitem{Singh1999}
R.~R.~P.~Singh, Z.~Weihong, C.~J.~Hamer, and J.~Oitmaa, Phys. Rev. B {\bf 60},
7278 (1999).

\bibitem{Tchernyshyov2003}
O.~Tchernyshyov, O.~A.~Starykh, R.~Moessner, A.~G.~Abanov, cond-mat/0301303
(2003).

\bibitem{Sachdev}
N.~Read and S.~Sachdev, Phys. Rev. Lett. {\bf 66}, 1773 (1991); \emph{ibid} {\bf
62}, 1694 (1989); G.~Murthy and S.~Sachdev, Nucl. Phys. B {\bf 344}, 557 (1990).

\bibitem{YoshimoriVillainLyonsKaplanNagamiya}
A.~Yoshimori, J.~Phys.~Soc.~Jpn. {\bf 14}, 807 (1959); J.~Villain,
J.~Phys.~Chem.~Solids {\bf 11}, 303 (1959); D.~H.~Lyons and T.~A.~Kaplan, Phys.
Rev. {\bf 120}, 1580 (1960).
%; T.~A.~Kaplan, \emph{ibid} {\bf 124}, 329 (1961).
T.~Nagamiya, in {\it Solid State Physics\/}, edited by F.~Seitz, D.~Turnbull and
H.~Ehrenreich, Vol.~20, 305 (Academic Press, New York, 1967); T.~Nagamiya,
T.~Nagata, and Y.~Kitano, Progr. Teor. Phys. {\bf 27}, 1253 (1962).

\bibitem{ZaliznyakZhitomirsky}
I. A. Zaliznyak, M. E. Zhitomirsky, cond-mat/0306370; JETP {\bf 81(3)}, 579
(1995); Phys. Rev. B {\bf 53}, 3428 (1996).

\bibitem{Zaliznyak2003}
I.~A.~Zaliznyak, unpublished (2003).

\bibitem{AndreevMarchenko}
A.~F.~Andreev and V.~I.~Marchenko, Sov.~Phys. Uspekhi {\bf 23}, 21 (1980).

\bibitem{Shender1982}
E.~Shender, Sov. Phys. JETP {\bf 56},178 (1982).

\bibitem{Henley1989}
C.~L.~Henley, Phys. Rev. Lett {\bf 62}, 2056 (1989); \emph{ibid} {\bf 73}, 2788
(1994).
\bibitem{Zaliznyak2000}
I.~A.~Zaliznyak, J.~P.~Hill, J.~M.~Tranquada, R.~Erwin, Y.~Moritomo,
Phys.~Rev.~Lett. {\bf 85}, 4353 (2000).

\end{thebibliography}
\end{document}